\documentclass[%
 aip,
 apl,
 amsmath,amssymb,
 reprint,%
]{revtex4-2}

\usepackage{graphicx}
\usepackage{dcolumn}
\usepackage{bm}

\usepackage[utf8]{inputenc}
\usepackage[T1]{fontenc}
\usepackage{mathptmx}
\usepackage{etoolbox}
\usepackage{siunitx}
\usepackage{pifont}

\makeatletter
\def\@email#1#2{%
 \endgroup
 \patchcmd{\titleblock@produce}
  {\frontmatter@RRAPformat}
  {\frontmatter@RRAPformat{\produce@RRAP{*#1\href{mailto:#2}{#2}}}\frontmatter@RRAPformat}
  {}{}
}%
\makeatother
\begin{document}

\preprint{AIP/123-QED}

\title[APL]{Mode-resolved spin-wave spectroscopy by a single paramagnetic center}
\author{C. Fourniols}
\affiliation{Université Grenoble Alpes, CEA, CNRS, Grenoble INP, Spintec, 38054 Grenoble, France}

\author{M. Umeda}
\affiliation{Advanced Science Research Center, Japan Atomic Energy Agency, Tokai, Ibaraki 319-1195, Japan}

\author{E. Clot}
\affiliation{Université Grenoble Alpes, CEA, CNRS, Grenoble INP, Spintec, 38054 Grenoble, France}

\author{E. Yildiz}
\affiliation{Université Grenoble Alpes, CEA, CNRS, Grenoble INP, Spintec, 38054 Grenoble, France}

\author{I. Joumard}
\affiliation{Université Grenoble Alpes, CEA, CNRS, Grenoble INP, Spintec, 38054 Grenoble, France}

\author{B. Pigeau}%
\affiliation{Université Grenoble Alpes, CNRS, Grenoble INP, Institut Néel, Grenoble, France}


\author{O. Klein}
 \email{oklein@cea.fr}
\affiliation{Université Grenoble Alpes, CEA, CNRS, Grenoble INP, Spintec, 38054 Grenoble, France}

\date{\today}
\begin{abstract}
We demonstrate mode-resolved ferromagnetic resonance (FMR) spectroscopy using optically detected magnetic resonance (ODMR) of a single nitrogen-vacancy (NV) center in (111)-oriented diamond. A two-tone drive couples the magnetization dynamics to the NV spin resonance via the magnon Kerr effect. We show that the NV can detect the static stray field produced by this effect down to precession cone angles $\theta \approx \SI{1}{\degree}$, providing a generic scheme to measure the linear FMR spectrum with the spatial resolution of an atomic-size probe. We validate the approach on a normally magnetized yttrium iron garnet (YIG) microdisk, resolving its intrinsic quantized spin-wave modes. Scanning the probe across the surface, we map the spatial profile of the modes and find quantitative agreement with analytical models.
\end{abstract}

\maketitle

\begin{figure}[h!]
\includegraphics[width=\columnwidth]{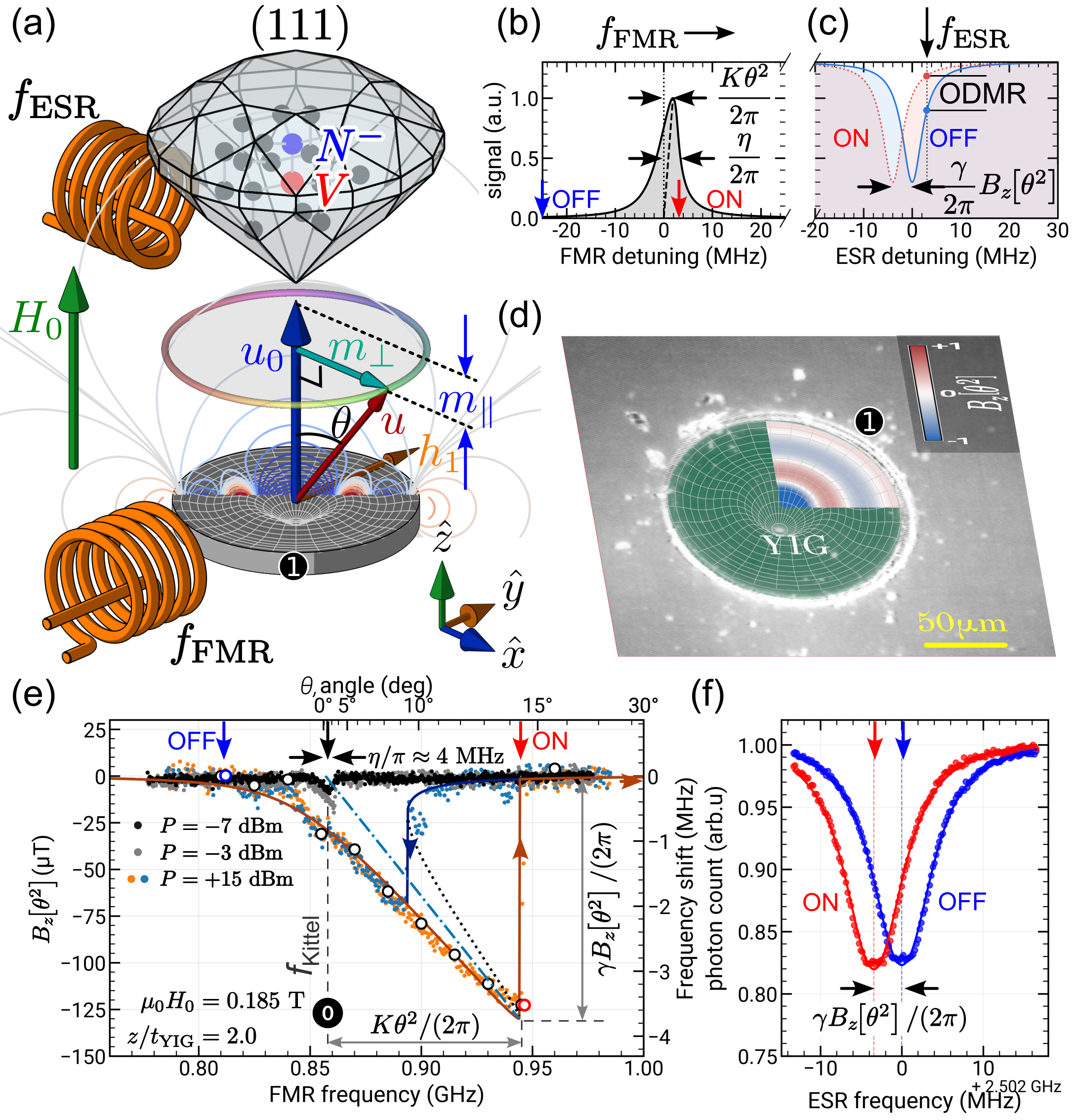}
\caption{\label{fig:1}  (a) Two-tone spectrometer: one microwave drive excites the FMR of a normally magnetized YIG micro-disk [image in (d)], the other the ESR of a single NV center with its quantization axis along the normal. (b) Magnon Kerr effect in the linear regime, $K \theta^2 < \eta$: tuning $f_\text{FMR}$ on or off the spin-wave resonance (red/blue) changes the static demagnetizing field $B_z[\theta^2]$ above the disk. (c) The resulting rigid shift of the ESR line is read out from the phase of the ODMR signal or from the ODMR amplitude at the inflection point of the ESR line. (e) ODMR phase signal versus FMR frequency near the Kittel mode: linear (black), bifurcation onset (grey) and foldover (orange/blue) regimes, with Duffing fits and Kerr shift (dash-dotted). (f) ESR line with the FMR detuned (OFF, blue) and at the jump point of an up-sweep (ON, red). The open circles in (e) are the ESR shifts measured as in (f).
}
\end{figure}

Optically detected magnetic resonance (ODMR) of nitrogen-vacancy (NV) centers has become a standard tool to image static~\cite{Casola2018,Huxter2022,Welter2023} and dynamic~\cite{Bertelli2020,Zhou2021,ManasValero2025,Finco2025} magnetic textures at the nanoscale. Applying it to ferromagnetic resonance (FMR)~\cite{Du2017,McCullian2020,Purser2020,LeeWong2020}, however, faces an obstacle. Producing a detectable change of ODMR at the FMR frequency with a single-tone microwave drive usually requires a field strength of the same order as the one used to drive the NV electron spin resonance (ESR), typically $\mu_0 h_1\approx \SI{100}{\micro\tesla}$. Such a level pushes the resonant dynamics of the ferromagnet well beyond its linear response, so that the spectral features reflect the strongly driven state instead of the intrinsic magnetic properties of the sample.

The issue is best viewed via the cone angle $\theta$, which measures the deviation of the magnetization from its local equilibrium. In steady state, $\theta$ reaches approximately $\gamma \mu_0 h_1/\eta$, where $\gamma$ is the gyromagnetic ratio and $\eta$ the half width at half maximum (HWHM) of the resonance. A distinctive feature of FMR is that the resonance frequency itself depends on $\theta$, $\omega(\theta) = \omega_0 + K\theta^2$: this is the magnon Kerr effect, which arises from the reduction of the longitudinal magnetization as the spins precess. At large power, it causes foldover and four-wave mixing (FWM)~\cite{AndersonSuhl1955,Goldberg1980,Seagle1985,Wang2016,Wang2018,Petrosyan2026}. The coefficient $K$ depends on the mode and sample geometry but is of order $\omega_M=\gamma\mu_0 M_s$, where $M_s$ is the saturation magnetization, i.e., a few GHz. This is a huge figure compared with $\eta$, which decomposes as $\eta_0 + \alpha \omega_0$, with $\eta_0$ the inhomogeneous broadening, $\alpha$ the Gilbert damping and $\omega_0/(2 \pi)$ the resonance frequency, and the sum is typically a few MHz. Mode-resolved spectroscopy, in which center frequency, linewidth and amplitude faithfully report the \emph{intrinsic} magnetic properties, requires the self-induced Kerr shift to stay below $\eta$, i.e., $\mu_0 h_1 \lesssim \sqrt{2 \eta^3/(\gamma^2\omega_M)}$. For low-damping materials such as yttrium iron garnet (YIG), with $\alpha \approx 10^{-4}$~\cite{Wolfe2014,Du2017}, this limits the drive at $\omega_0=(2\pi)\SI{1}{GHz}$ to $\mu_0 h_1 \lesssim \SI{1}{\micro\tesla}$, i.e., $\theta\approx\SI{1}{\degree}$. ESR drive levels are about two orders of magnitude larger, pushing the magnetization dynamics in the regime dominated by hysteretic linewidth broadening~\cite{Wang2023} and FWM among magnon modes~\cite{McCullian2020}.

In this letter, we demonstrate that the magnon Kerr effect, despite being a second-order effect in $\theta$, allows the detection of the \emph{linear} FMR spectrum with a single NV center. Our approach rests on two ingredients: \textit{i)} a two-tone scheme that decouples the strong ESR drive (of order \SI{100}{\micro\tesla}) from a weak FMR drive (a few \si{\micro\tesla}), calibrated in situ from the onset of foldover bistability; \textit{ii)} the detection of the static stray field produced by the reduction of the longitudinal magnetization through the ODMR of a single NV center. We show that the NV resolves this field down to $\theta\approx\SI{1}{\degree}$, i.e., $K \theta^2/2\pi \approx 1$~MHz, far below previously reported values~\cite{VanderSar2015,Merbouche2022,Lake2022,McCullian2025}, and hence within the linear regime. Because every precessing magnet reduces its longitudinal magnetization, the scheme depends neither on the material nor on the mode, and it requires no frequency matching with the NV transition.

We validate the method on a normally magnetized YIG micro-disk and obtain three main results: \textit{i)} the linear spectrum of the quantized spin-wave modes is resolved with a HWHM of 2~MHz, and the mode frequencies and their field dependence agree with a semi-analytical model without adjustable parameters; \textit{ii)} comparing the internal (Kerr) and external (ESR) shifts quantifies $\theta$; \textit{iii)} scanning the NV probe across the disk maps the spatial profile of the modes. We show that the position of the sign reversal of the ODMR signal is compatible with a precession profile pinned at the disk edge.

Figure~\ref{fig:1}(a) provides a schematic of the experimental setup. The ODMR sensor is a $(111)$-oriented micro-diamond hosting a single NV$^-$ center with quantization axis along $\hat z$, attached at the apex of a cantilever glued onto a tuning fork, which a three-axis actuator translates relative to the sample. As reference sample, we chose a disk of radius $R_\text{YIG}=\SI{84}{\micro\meter}$ and thickness $t_\text{YIG}=\SI{4.75}{\micro\meter}$, extensively characterized by inductive detection and by magnetic resonance force microscopy~\cite{Charbois2002JAP,Charbois2002APL,Naletov2003}. Its spectrum is dominated by the magnetic dipolar interaction and hence fully described by a Walker-type analytical model~\cite{Walker1957} (see below), and its aspect ratio $R_\text{YIG}/t_\text{YIG} \approx 18$ ensures a large free spectral range between bright modes sharing the same zero orbital angular momentum of the antenna. The disk was patterned by dry ion milling from a $(111)$ YIG film grown by liquid-phase epitaxy on gadolinium gallium garnet. Figure~\ref{fig:1}(d) shows an image of the disk, whose magnetic parameters are summarized in Table~\ref{tab:param}.

\begin{table}
  \caption{Magnetic parameters of the YIG disk: thickness, radius, saturation magnetization, inhomogeneous broadening, Gilbert damping and gyromagnetic ratio.}
  \label{tab:param}
  \begin{ruledtabular}
    \begin{tabular}{c c c c c c}
      $t_\text{YIG}$  & $R_\text{YIG}$ & $\mu_0 M_s$ & $\eta_0/(2\pi)$  &  $\alpha$  &  $\gamma/(2 \pi)$  \\
     (\si{\micro\meter})  &  (\si{\micro\meter}) & (T) &  (MHz) &  & (\si{GHz/T}) \\ \hline
      4.75 & 84 & 0.17 & 2 & $10^{-4}$ & $28.7$  \\
    \end{tabular}
  \end{ruledtabular}
\end{table}

The disk is uniformly magnetized by an external field $H_0 > M_s$ applied along its normal and produced by a permanent magnet placed underneath the sample. Field scans are performed by translating the magnet axially. The field is aligned with the disk normal by rotating the magnet relative to the sample to minimize the Kittel frequency at fixed field amplitude.

As illustrated in Fig.~\ref{fig:1}(a), the magnetization dynamics of the disk decomposes into a transverse component ${\bm m}_\perp \propto \theta$, which produces an oscillating stray field, and a longitudinal component ${\bm m}_\parallel \propto \theta^2$, which produces a static one and is responsible for the Kerr effect. We define $B_{z}\left[ \theta^2 \right](\bm{x})$ as the static stray magnetic field produced by $m_\parallel = \omega_M \left ( 1 - \cos \theta \right ) / \gamma$. The cone angle $\theta$ decreases from the center to the edge of the disk, $\theta(r) = \theta_0\,\psi(r)$, where $\theta_0$ is the cone angle at the center and $\psi(r)$ a bell-shaped profile with $\psi(0) = 1$. For thin disks, neglecting the thickness dependence of the precession profile, a simple analytical expression can be derived:
\begin{equation}
  \begin{aligned}
    B_{z}\left[\theta^2\right](\bm{x})
    &\approx
    -\frac{1}{2}\frac{\omega_M}{\gamma}\,\theta_0^2
    \biggl[\int_0^{R_\text{YIG}}
      \partial_r \psi^2(r)\,
      {N}_{zz}\left[r,t_\text{YIG}\right](\bm{x})\,dr \\
    &\qquad
    + \psi^2(R_\text{YIG})\,
      {N}_{zz}\left[R_\text{YIG},t_\text{YIG}\right](\bm{x})
    \biggr].
  \end{aligned}
  \label{eq:dzz}
\end{equation}
Here $\mu_0 M_s {N}_{zz}\left[ r, t_\text{YIG} \right](\bm x)$ is the demagnetizing field along $z$ at position $\bm x$ produced by a cylinder of radius $r$ and thickness $t_\text{YIG}$ uniformly magnetized along $z$. An analytical expression for the demagnetization tensor $\overline{\overline{N}}$  was first derived by L.~Kraus~\cite{Kraus1973}. Throughout, $\partial_r$ denotes the partial derivative with respect to the radial position $r$.

Figures~\ref{fig:1}(b) and (c) illustrate the measurement scheme. The experiment consists of monitoring either i) the phase of the photoluminescence signal relative to a frequency-modulated ESR drive~\cite{Welter2023} or ii) the photoluminescence amplitude of the NV center under continuous-wave excitation at a fixed $f_\text{ESR}$ set at the inflection point on the left wing of its absorption curve, while scanning the second tone around $f_\text{FMR}$ at much lower power.

We first focus on the spectroscopy with the NV center on the disk axis, at a distance $2 t_\text{YIG}$ above the surface and in a bias field $\mu_0 H_0=0.185$~T. Figure~\ref{fig:1}(e) shows the static stray magnetic field produced by the Kittel mode (extracted from the ODMR phase signal) as a function of the FMR frequency for different drive powers. At high power ($+15$~dBm), the response depends on the sweep direction, the hallmark of foldover~\cite{AndersonSuhl1955,McKinstry1985}. The data are well fitted (solid lines) by a Duffing curve, $\left( ({\omega_0 - \omega})/{\eta} + \beta y \right)^2 + 1  = 1/y$, where $y$ is the normalized signal, proportional to $\theta^2$, and $\beta = \frac{1}{2}\omega_M \left(\gamma \mu_0 h_1\right)^2/\eta^3$ measures the nonlinear frequency shift in units of $\eta$. The fit yields $\beta = 8.7$ and $\eta^\ast/(2 \pi) = 9.2$~MHz\footnote{Here $\beta$ and $\eta^\ast$, which replaces $\eta$, are free fit parameters, not computed from $h_1$. The broadening relative to $\eta$ is attributed to FWM~\cite{sparks}.}. For such large $\beta$, two stable branches (solid lines), separated by an unstable one (dotted line), coexist over a finite frequency range, and the up- and down-sweeps jump at its upper and lower edges, respectively. Along the up-sweep, the signal increases linearly with frequency, as expected for a shift proportional to $\theta^2$: this confirms that the magnon Kerr effect is the leading nonlinearity. The dash-dotted line shows the corresponding Kerr shift $\omega(\theta) - \omega_0$ of the Duffing resonance. Figure~\ref{fig:1}(f) shows the ESR line with the FMR detuned (OFF, blue) and at the jump point of the up-sweep (ON, red): the phase signal arises from a pure rigid shift of the ESR line, without the lineshape distortions reported in other studies\footnote{We do not observe any increase of the linewidth as reported in previous studies~\cite{Du2017}.}. The open circles, obtained by fitting the ESR center frequency at different FMR detunings, confirm that the phase signal matches the spectral shift.

Reducing the power reduces $\beta$. When $\beta$ reaches the critical value $\beta_c = 8/(3\sqrt{3})$, the stable and unstable branches merge in a saddle-node bifurcation, where $\partial_\omega \theta \to \infty$, which experimentally marks the onset of hysteresis. In our experiment, this occurs at $P=-3$~dBm (grey dots). When the power is further reduced to $P=-7$~dBm (black dots), the FMR line is still detected, demonstrating that the Kerr coupling allows access to the linear response. A Lorentzian fit of the linear peak yields $\eta/(2 \pi) = 2$~MHz. This HWHM is in line with previous studies~\cite{Charbois2002} of the inhomogeneous broadening of the same disk (see Table~\ref{tab:param}). We use the onset of bistability to calibrate $h_{1}$ at the disk position: at $P=-3$~dBm, we find $\mu_0 h_{1, \text{FMR}} = \SI{2.4}{\micro\tesla}$. By comparison, a field of $\mu_0 h_{1, \text{ESR}} = \SI{0.1}{\milli\tesla}$ is used to drive the ESR.

We now estimate the precession angle. We can either infer $\theta$ from the change of the internal dipolar field, measured by the nonlinear frequency shift, or from the change of the external dipolar field, measured by the rigid shift of the ESR line. Both depend on the normalized precession profile $\psi(r)$ introduced above. For such a non-uniform profile, we define
\begin{equation}
  \{\theta\} = \theta_0 \, \frac{\int_0^{R_\text{YIG}} m_\parallel(r)\,\psi(r)\,r\,dr}{\int_0^{R_\text{YIG}} m_\parallel(r)\,r\,dr}  , 
  \label{eq:thetaavg}
\end{equation}
the spatial average of $\psi$ weighted by the profile of $m_\parallel$, which itself depends on $\theta$. The nonlinear frequency shift then takes the simple form
\begin{equation}
\omega(\theta) \approx  \omega_0 + \dfrac 1 2 \omega_M \{\theta\}^2.
\end{equation}

\begin{figure}[t!]
\includegraphics[width=\columnwidth]{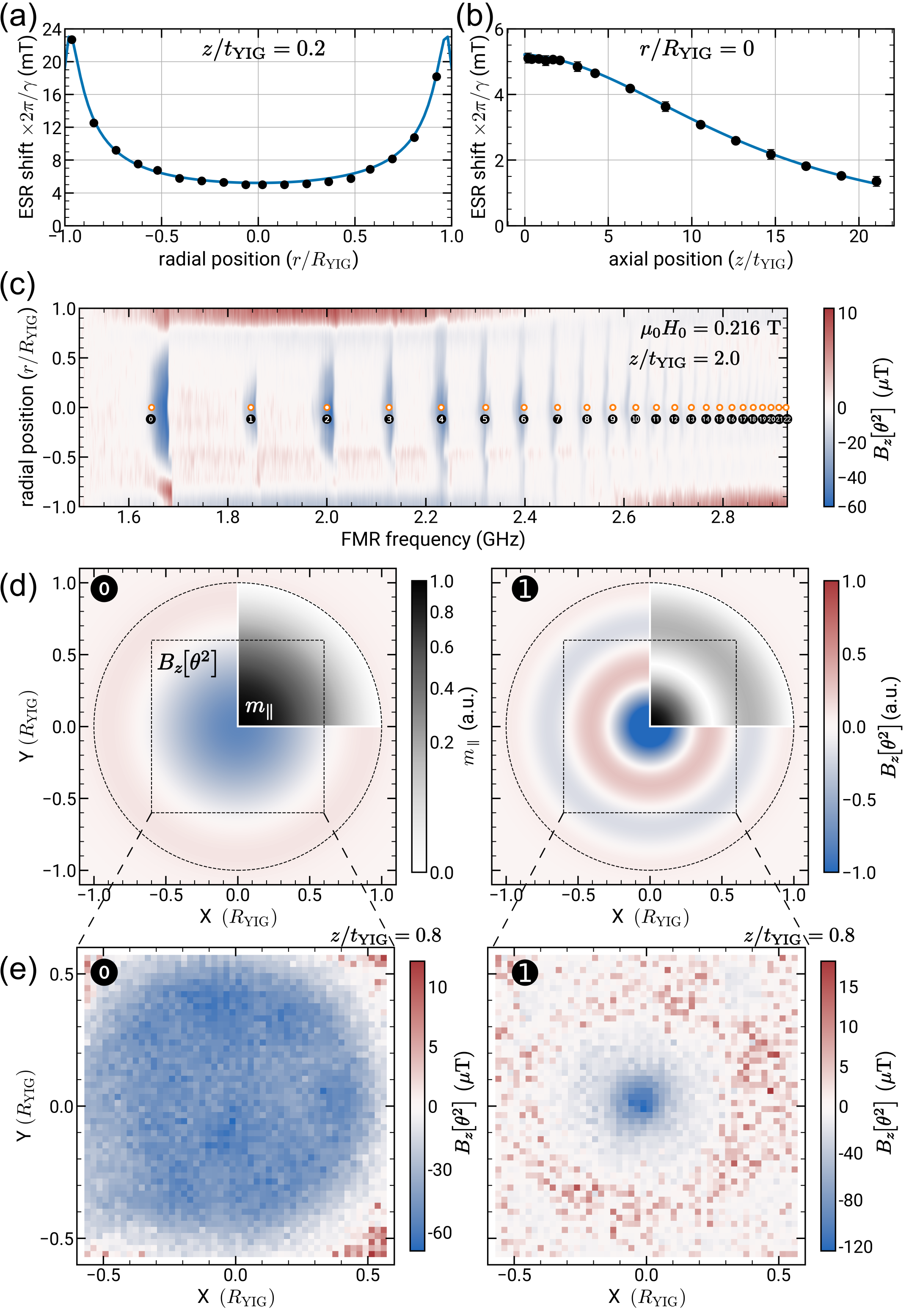}
\caption{\label{fig:3} (a), (b) Static stray field of the YIG disk, measured from the ESR shift relative to far from the disk, versus (a) radial and (b) axial position of the diamond at $\mu_0 H_0 = 0.216$~T. (c) Evolution of the ODMR-measured FMR spectra versus $r/R_\text{YIG}$ (up-scan, foldover). The power is gradually increased to compensate weaker coupling with higher order radial modes. The orange dots are the analytical prediction of the eigen-frequency of the radial modes. (d) Predicted $B_z\left [ \theta^2 \right ]$ for the Kittel mode and its first higher radial mode. (e) Experimental observation in the region $\left [ -0.6, +0.6\right ] R_\text{YIG}$. The measurement is performed at $\mu_0 H_0 = 0.280$~T.}
\end{figure}

Assuming a profile $\psi(r) = J_0(\kappa_0 \,r/R_\text{YIG})$, where $J_0$ is the zeroth-order Bessel function and  $\kappa_{0} = 2.405$ is its first root (corresponding to a total pinning boundary condition at the periphery, $\psi(R_\text{YIG}) = 0$~\cite{Kakazei2004,Guslienko2005}), we obtain $\{\theta\} = 0.723\,\theta_0$. Hence the 80~MHz Kerr shift observed in Fig.~\ref{fig:1}(e) at $+15$~dBm corresponds to a precession angle $\theta_0 = \SI{14}{\degree}$ at the disk center. We use this conversion to set the precession-angle scale on the upper abscissa. For the same power and profile, with the NV on the disk axis at $2 t_\text{YIG}$ above the surface, Eq.~(\ref{eq:dzz}) predicts $\gamma B_z[\theta^2] = 0.053 \omega_M \theta_0^2/2$; the measured ESR shift of 4~MHz then yields $\theta_0 = \SI{10}{\degree}$. The 30\% discrepancy between the two estimates is likely due to the change of the normalized profile $\psi(r)$ with power, which gradually flattens near the center, as visible in Fig.~\ref{fig:3}(e). This improves the match by increasing $\{\theta\} / \theta_0$. Another possible contribution is the thickness dependence of the precession profile, neglected in Eq.~(\ref{eq:dzz}): including it reduces $B_z[\theta^2] / \theta_0^2$ and further improves the match.

\begin{figure}[t!]
\includegraphics[width=\columnwidth]{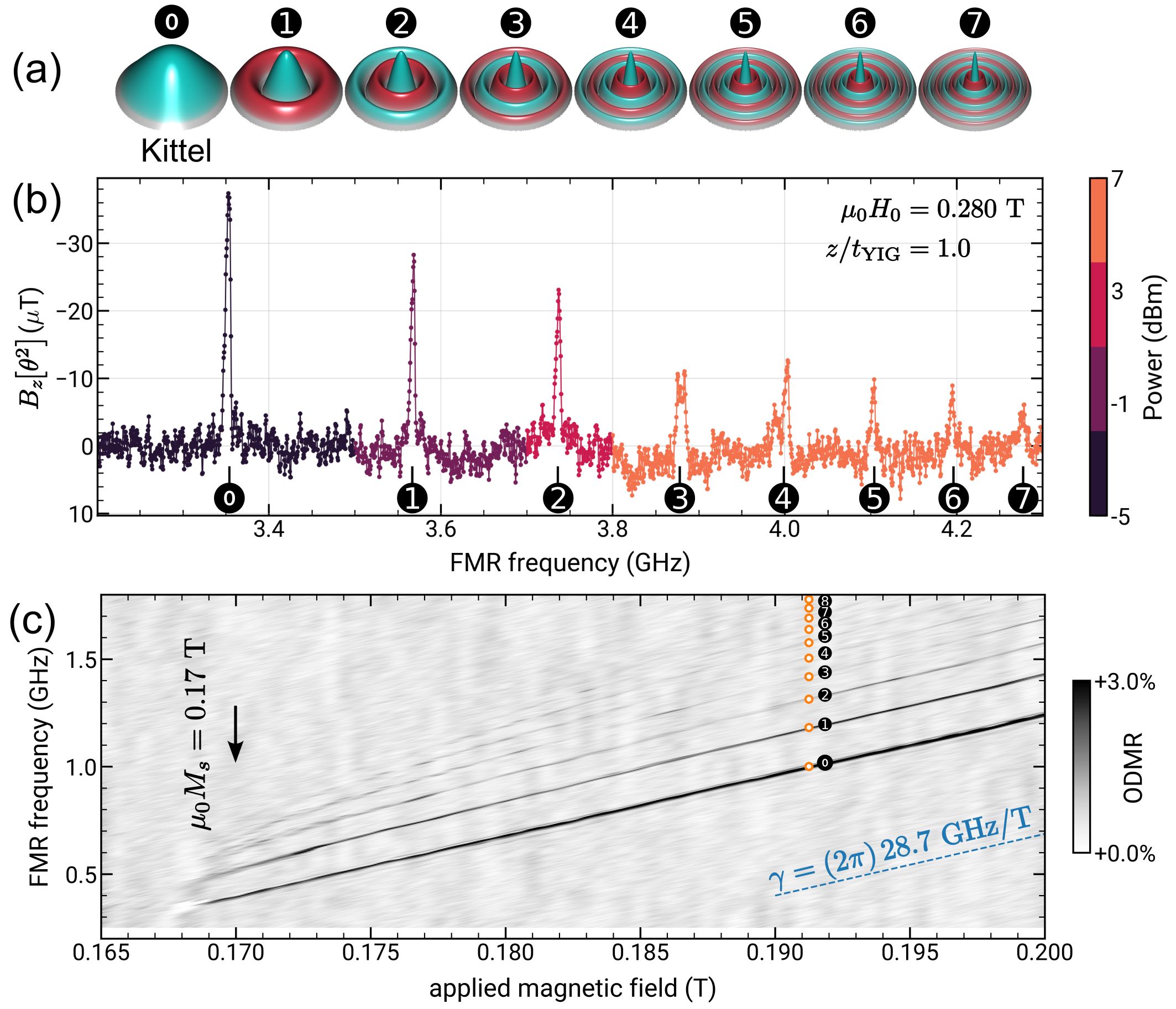}
\caption{\label{fig:2} (a) Spatio-temporal profiles of the quantized spin-wave modes with zero orbital angular momentum in a saturated, normally magnetized disk; color encodes the phase (color wheel as in Fig.~\ref{fig:1}(a); complementary colors mark antiphase precession). (b) ODMR-detected linear spectrum at $z = 2 t_\text{YIG}$, $r = 0$ and $\mu_0 H_0 = 0.280$~T. The drive power is ramped up during the sweep to offset the weaker antenna coupling of higher radial modes; ticks mark their predicted positions. (c) Measured frequency-field dispersion of the magnon spectrum at $P = 0$~dBm. The orange dots are the analytical prediction of the eigen-frequency of the radial modes at $H_0=0.1911$~T (see text).}
\end{figure}

Having established that the Kerr coupling gives access to the linear response, we show in Fig.~\ref{fig:2}(b) the full linear spectrum, measured at $\mu_0 H_0 = 0.280$~T with the diamond at the same position as in Fig.~\ref{fig:1}(e). The uniform microwave field of the antenna selectively excites modes with zero orbital angular momentum~\cite{Valet2026-1,Valet2026-2}, whose precession profiles are shown in Fig.~\ref{fig:2}(a). The most intense peak is the Kittel mode, followed by the higher-order radial modes, labeled by their number of radial nodes.

To identify the measured resonances, we compute the spin-wave spectrum of the disk semi-analytically. Linearized around the saturated state, the Landau-Lifshitz-Gilbert (LLG) equation for the circular amplitude takes the form of a stationary Schr\"odinger equation, in which the internal field plays the role of the potential and the exchange and dipolar interactions that of the kinetic term. In a normally magnetized disk, the internal field $H_i(r) = H_0 - M_s \overline{N}_{zz}(r)$, where $\overline{N}_{zz}(r)$ is the thickness average of the depolarization tensor of Eq.~(\ref{eq:dzz}), is minimal at the center and rises steeply near the edge [Fig.~\ref{fig:3}(a)]: the disk acts as a cylindrical potential well for spin waves. Following Damon and Van de Vaart~\cite{Damon1965,Damon1965a}, we treat the in-plane dynamics locally: at each radius, spin waves obey the dispersion relation of the lowest thickness mode of an infinite film of thickness $t_\text{YIG}$ placed in the local field $H_i(r)$, $p^{1/2}\tan\left(p^{1/2} k\,t_\text{YIG}/2\right) = 1$, with $p = \left[(B_i/\mu_0) H_i - (\omega/\gamma\mu_0)^2\right]/\left[(\omega/\gamma\mu_0)^2 - H_i^2\right]$ and $B_i/\mu_0 = H_i + M_s$~\cite{Damon1965}. This yields a local wavevector $k(r,\omega)$, and the radial modes with zero azimuthal index follow from the Wentzel-Kramers-Brillouin (WKB) condition
\begin{equation}
  \int_0^{r_t(\omega)} k(r,\omega)\,dr = \pi\left(n_R + \dfrac 1 2 \right),
  \qquad n_R = 0, 1, 2, \dots
  \label{eq:wkb}
\end{equation}
where $r_t$ is the turning point, $\gamma\mu_0 H_i(r_t) = \omega$, beyond which waves are evanescent, and the term $1/2$ accounts for the phases acquired at the two ends of the radial path, the axis and the turning point.\footnote{Each end contributes $1/4$. On the axis, a wave traveling along a diameter acquires a Gouy phase $\pi/2$ as it crosses the focal line, half of which is assigned to each radial half-path; equivalently, the regular solution behaves as $J_0(kr) \propto \cos(kr - \pi/4)$. At a smooth turning point, the Airy connection gives the standard phase $\pi/4$. Eq.~(\ref{eq:wkb}) is then the Langer-corrected Bohr-Sommerfeld rule~\cite{Langer1937}, which is exact for the two-dimensional harmonic oscillator. A hard wall with pinned boundary condition would instead contribute $1/2$, i.e., $3/4$ in total; the steep but finite field rise at the rim lies between these two limits, a difference that becomes negligible for large $n_R$.} The ticks in Fig.~\ref{fig:2}(b) are the solutions of Eq.~(\ref{eq:wkb}), with no adjustable parameter beyond those of Table~\ref{tab:param}.

The model neglects exchange, retains only the lowest thickness mode, and assumes that $H_i$ varies slowly on the scale of the wavelength, which fails near the edge and for the lowest modes. The agreement nevertheless extends to all $n_R$: for large $n_R$, WKB becomes asymptotically exact, while for $n_R = 0$ the frequency is pinned just above the bottom of the well, $\gamma\mu_0 H_i(0)$, which is wide and flat owing to the large aspect ratio.

Figure~\ref{fig:2}(c) shows the evolution of the spectrum with the applied field. The Kittel mode depends linearly on $H_0$ down to saturation, which indicates that the applied field is aligned with the normal to within \SI{0.1}{\degree}. At high field, the whole spectrum translates rigidly with $H_0$, following the gyromagnetic ratio (blue dashed line), whereas it compresses as the field is reduced. The Kittel mode cannot depart far from the bottom of the well, $\gamma\mu_0 H_i(0)$, which shifts with $H_0$ at the rate $\gamma\mu_0$. Higher radial modes accumulate below $\gamma\mu_0 \sqrt{H_i(0)\left[H_i(0) + M_s\right]}$, reached for $k \to \infty$ at the center. The width of this window tends to $\omega_M/2$ at high field, but closes as $\gamma\mu_0 \sqrt{M_s H_i(0)}$ for $H_i(0) \ll M_s$, which explains the compression at low field. Resonances above this bound can only propagate in an annulus near the rim, where the internal field is larger.

We finally study in Fig.~\ref{fig:3} how the spectra vary as the probe is displaced. Figures~\ref{fig:3}(a) and (b) show the ESR shift, relative to its value far from the disk, as a function of the radial and axial position of the diamond tip. As expected, this shift, which reflects the static stray field of the saturated disk, depends strongly on the tip position. The solid lines are fits with the analytical expression of $\mu_0 M_s {N}_{zz}[R_\text{YIG}, t_\text{YIG}] (\bm x)$~\cite{Kraus1973}.

Next, Fig.~\ref{fig:3}(c) shows the corresponding variation of the spectral fingerprint as a function of the radial position of the diamond, measured up to the 20th radial mode by gradually increasing the power from $P=7$ to $22$~dBm during an upward frequency sweep. The asymmetric line shape is due to foldover, as described in Fig.~\ref{fig:1}(e). In sharp contrast to the ESR shift, the frequency of the foldover jump hardly depends on the tip position, reflecting the collective nature of the spin-wave modes. The orange dots mark the eigenfrequencies predicted by Eq.~(\ref{eq:wkb}) and they reproduce well the expected positions of the spectral lines.

We have also performed a full mapping of the precession profile as the diamond tip is scanned horizontally above the surface. Figure~\ref{fig:3}(d) shows the expected $B_z [ \theta^2 ]$ map for $m_\parallel$ profiles following a Bessel shape with total pinning at the edge, $\psi(r) = J_0(\kappa_n r/R_\text{YIG})$, where $\kappa_n$ is the $(n+1)$th root of $J_0$. Figure~\ref{fig:3}(e) shows the experimental result \footnote{The phase signal is converted into a field with a ratio of \SI{1.5}{\micro\tesla/\degree}}. For the Kittel mode, the ODMR signal inverts its polarity at $r/R_\text{YIG} \approx 0.6$ as predicted in (d). Since this radial position of the polarity inversion of the ODMR signal depends on the pinning conditions at the boundary, our measurement of the Kittel mode static stray field is compatible with a precession profile that vanishes at the periphery of the disk.

In summary, we have performed mode-resolved spectroscopy of the quantized spin-wave modes of a normally magnetized YIG micro-disk using a single NV center. Single-NV ODMR detects the change of the static stray field produced by the magnon Kerr effect even in the linear regime, resolving linewidths of a few MHz, and the measured mode frequencies agree with a WKB model of the dipolar spectrum without adjustable parameters. Comparing the Kerr shift with the ESR shift calibrates the precession angle, and scanning the diamond across the disk maps the radial precession profile, which suggests that the Kittel mode is pinned at the disk edge. Altogether, these results establish ODMR as a quantitative, MHz-resolution probe of the intrinsic spin-wave spectrum of magnetic microstructures.

\begin{acknowledgments}
 We sincerely thank K. Yamamoto from the Japan Atomic Energy Agency (JAEA) for his invaluable insights and thought-provoking discussions. This work was partially supported by the PEPR SPIN - MAGISTRAL ANR-24-EXSP-0004; the Program QuanTEdu-France n° ANR-22-CMAS-0001 France 2030, the CEA PTC program 26MP33 TOPGUN, the French Minatech program, and the REIMEI Research Program of JAEA. 
 
\end{acknowledgments}

\section*{Data Availability}
The data that support the findings of this study are available from the corresponding author upon reasonable request.


%

\end{document}